\documentclass[aps,prl,twocolumn,preprintnumbers,amsmath,amssymb,superscriptaddress]{revtex4} 
\usepackage{graphicx}
\usepackage{bm}
\usepackage[final]{pdfpages}

\begin{document}
\title{Evidence of Majorana Zero Modes in Josephson Trijunctions}
\date{\today} 

\author{Guang Yang}\thanks{These authors contribute equally to this work.}
\affiliation{Beijing National Laboratory for Condensed Matter Physics, Institute of Physics, Chinese Academy of Sciences, Beijing 100190, China} \affiliation{School of Physical Sciences, University of Chinese Academy of Sciences, Beijing 100049, China}
\author{Zhaozheng Lyu}\thanks{These authors contribute equally to this work.}
\affiliation{Beijing National Laboratory for Condensed Matter Physics, Institute of Physics, Chinese Academy of Sciences, Beijing 100190, China} \affiliation{School of Physical Sciences, University of Chinese Academy of Sciences, Beijing 100049, China}
\author{Junhua Wang}
\affiliation{Beijing National Laboratory for Condensed Matter Physics, Institute of Physics, Chinese Academy of Sciences, Beijing 100190, China} \affiliation{School of Physical Sciences, University of Chinese Academy of Sciences, Beijing 100049, China}
\author{Jianghua Ying}
\affiliation{Beijing National Laboratory for Condensed Matter Physics, Institute of Physics, Chinese Academy of Sciences, Beijing 100190, China} \affiliation{School of Physical Sciences, University of Chinese Academy of Sciences, Beijing 100049, China}
\author{Xiang Zhang}
\affiliation{Beijing National Laboratory for Condensed Matter Physics, Institute of Physics, Chinese Academy of Sciences, Beijing 100190, China} \affiliation{School of Physical Sciences, University of Chinese Academy of Sciences, Beijing 100049, China}
\author{Jie Shen}\thanks{Present address: QuTech and Kavli Institute of Nanoscience, Delft University of Technology, 2600 GA Delft, The Netherlands.} \affiliation{Beijing National Laboratory for Condensed Matter Physics, Institute of Physics, Chinese Academy of Sciences, Beijing 100190, China}
\author{Guangtong Liu}
\affiliation{Beijing National Laboratory for Condensed Matter Physics, Institute of Physics, Chinese Academy of Sciences, Beijing 100190, China}
\author{Jie Fan}
\affiliation{Beijing National Laboratory for Condensed Matter Physics, Institute of Physics, Chinese Academy of Sciences, Beijing 100190, China}
\author{Zhongqing Ji}
\affiliation{Beijing National Laboratory for Condensed Matter Physics, Institute of Physics, Chinese Academy of Sciences, Beijing 100190, China}
\author{Xiunian Jing}
\affiliation{Beijing National Laboratory for Condensed Matter Physics, Institute of Physics, Chinese Academy of Sciences, Beijing 100190, China}
\author{Fanming Qu}\email[Corresponding authors: ]{fanmingqu@iphy.ac.cn}
\affiliation{Beijing National Laboratory for Condensed Matter Physics, Institute of Physics, Chinese Academy of Sciences, Beijing 100190, China} \affiliation{CAS Center for Excellence in Topological Quantum Computation, University of Chinese Academy of Sciences, Beijing 100190, China}
\author{Li Lu} \email[Corresponding authors: ]{lilu@iphy.ac.cn}
\affiliation{Beijing National Laboratory for Condensed Matter Physics, Institute of Physics, Chinese Academy of Sciences, Beijing 100190, China} \affiliation{School of Physical Sciences, University of Chinese Academy of Sciences, Beijing 100049, China} \affiliation{CAS Center for Excellence in Topological Quantum Computation, University of Chinese Academy of Sciences, Beijing 100190, China}


\begin{abstract}
In search of fault-tolerant topological quantum computation (TQC), zero-bias conductance peak as a necessary signature of Majorana zero modes (MZMs) has been observed in a number of solid-state systems. Here, we present the signature of MZMs from a phase-sensitive experiment on Josephson trijunctions constructed on the surface of three-dimensional topological insulators. We observed that the minigap at the center of the trijunction is protected to close over extended regions in phase space, supporting in principle the Majorana phase diagram proposed by Fu and Kane in 2008. Our study paves the way for further braiding MZMs and exploring TQC on a scalable two-dimensional platform.
\end{abstract}

\maketitle

It is believed that fault-tolerant TQC can be realized by encoding quantum information on topologically protected quantum states \cite{1,2,3}. In 2001, Kitaev proposed the use of p-wave superconducting chains to host MZMs as topological qubits \cite{4}. In 2008, Fu and Kane further proposed to induce p-wave-like superconductivity from s-wave superconductors via proximity effect in a hybrid structure \cite{5}. Since then, many hybrid structures have been proposed \cite{6,7,8,9,10}, and signatures of MZMs have been observed in structures containing semiconducting nanowires \cite{11,12,13,14,15}, topological insulators \cite{16,17,18,19,20}, iron chains \cite{22}, etc. However, the original proposal of Fu and Kane -- to construct Josephson trijunctions on topological insulators \cite{5}, which could potentially serve as the basic components for universal TQC \cite{23,24} -- remains unexplored.

\begin{figure}
\includegraphics[width=1 \linewidth]{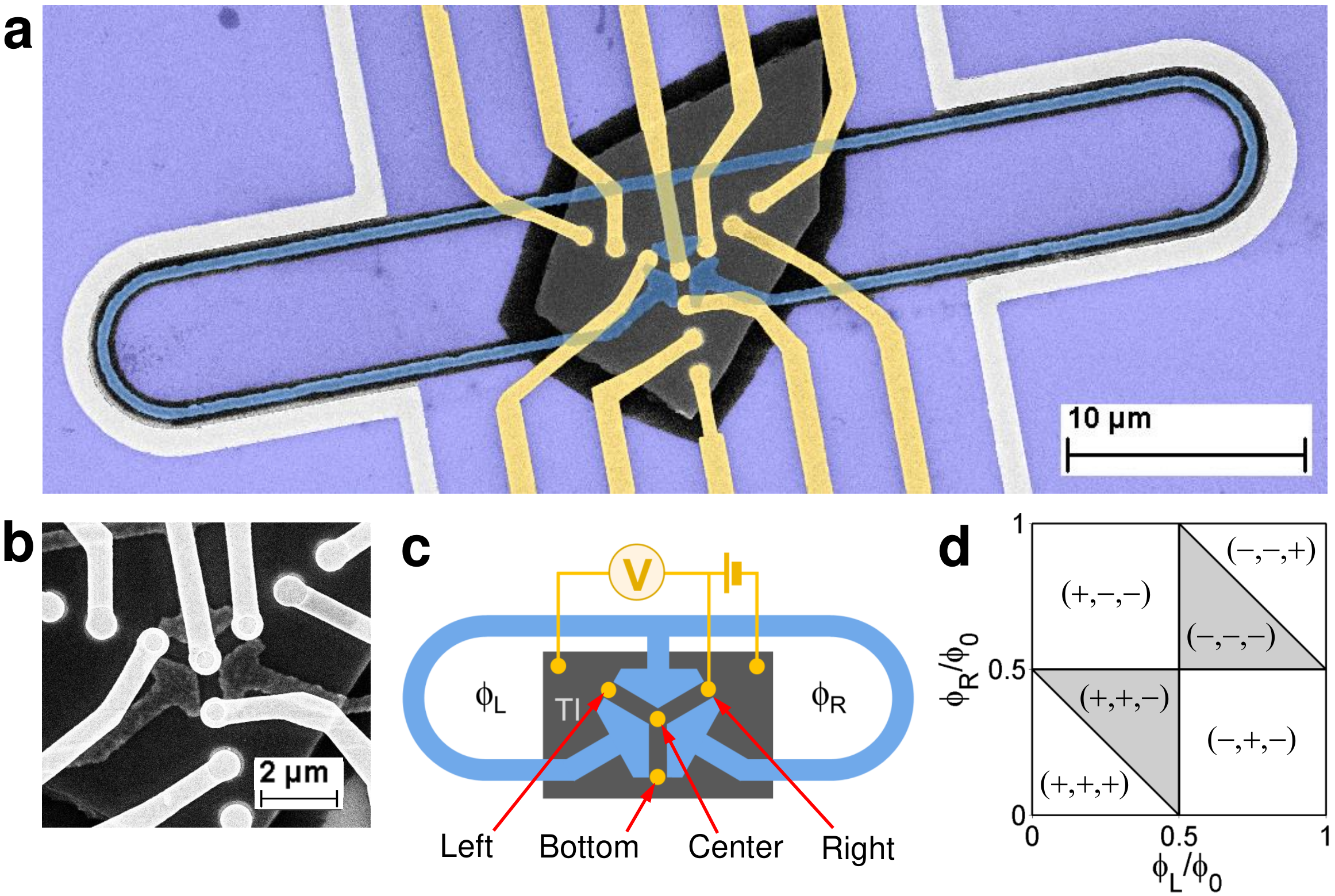}
\caption{\label{fig:fig1} {(\textbf{a}) False-color SEM image of the 1$^{\rm st}$ device. A Pb Josephson trijunction (in blue) was fabricated on the surface of a Bi$_2$Te$_3$ flake. By applying electric currents to the two Al or Nb half-turn coils (in silvery), the magnetic flux in the loops, thus the phase difference in corresponding junctions, can be adjusted independently. For detecting the local ABSs, normal-metal Au electrodes (in yellow) were fabricated to contact with the Bi$_2$Te$_3$ surface at the center and the ends of the trijunction, through holes on the blackish-looking insulating mask. (\textbf{b}) The central part of the device. (\textbf{c}) Schematic of the trijunction device and the three-terminal configuration for contact resistance measurement. (\textbf{d}) Fu-Kane's MZM phase diagram \cite{5} for the center of the trijunction in the loop-flux space. MZMs are expected in the shadowed regions where minigap-inversion occurs in odd numbers of single junctions (shown in the brackets are the signs of the minigap in the left, the right, and the bottom single junctions). }}
\end{figure}

According to Fu and Kane \cite{5}, for Josephson trijunctions constructed on the surface of a three-dimensional topological insulator (3D TI), there will be a boundary at the center isolating the single junctions with positive minigap to those with negative minigap, when gap-inversion occurs in odd numbers of single junctions. Such a boundary, at which the minigap closes completely and a localized MZM appears, is protected to occur over extended parametric regions with nontrivial topological numbers, as illustrated in Fig. 1d. The verification of complete minigap-closing over extended regions in phase space, in analogy to various quantum Hall edge states surviving over extended parametric regions, would provide strong evidence for the existence of MZMs in TI-based Josephson devices.

\begin{figure*}
\includegraphics[width=0.90 \linewidth]{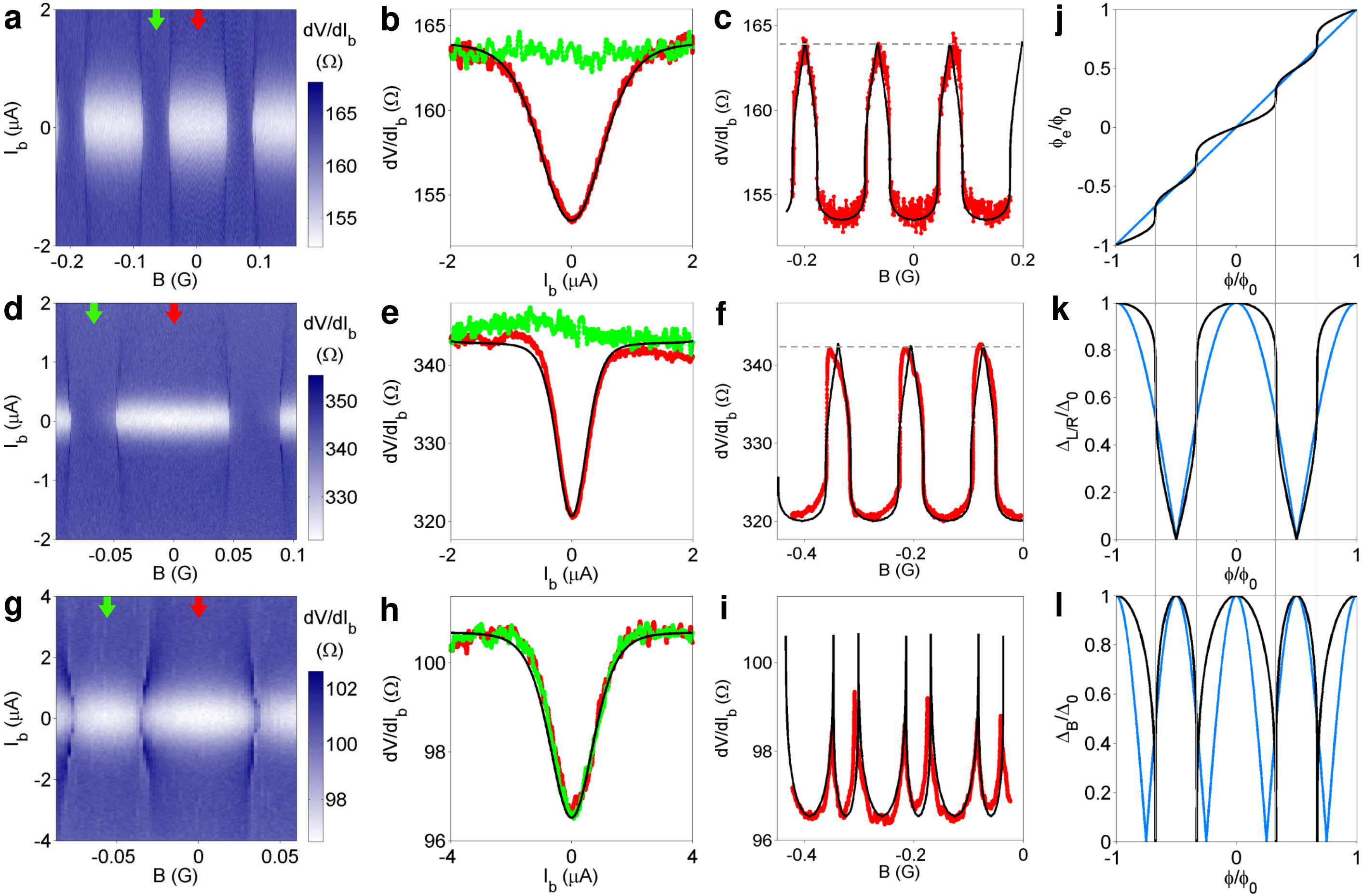}
\caption{\label{fig:fig2} {The contact resistance $dV/dI_{\rm b}$ measured at the ends of the 1$^{\rm st}$ trijunction at $T$=0.5 K. (\textbf{a}) The $dV/dI_{\rm b}$ measured at the left end, as functions of global magnetic field $B$ and bias current $I_{\rm b}$. (\textbf{b}) The vertical line cuts of (a) at magnetic fields indicated by the arrows in corresponding color in (a). The data are fitted by using the BTK theory (black line). (\textbf{c}) The horizontal line cut of (a) at $I_{\rm b}$=0. (\textbf{d}), (\textbf{e}), (\textbf{f}) and (\textbf{g}), (\textbf{h}), (\textbf{i}) Similar data measured at the right end and the bottom end of the trijunction, respectively. (\textbf{j}) Effective magnetic flux $\phi_{\rm e}$ in the superconducting loop as a function of applied magnetic flux $\phi$, when the screening parameter of the loop $\beta$=0 (blue line) and $\beta$=0.5 (black line). (\textbf{k}) and (\textbf{l}) The theoretical flux dependences of the minigap in the left/right junctions (k) and in the bottom junction (l), when $\beta$=0 (blue lines) and $\beta$=0.5 (black lines). By using the functional forms of these black lines, the magnetic field dependences of $dV/dI_{\rm b}$ in (\textbf{c}), (\textbf{f}), (\textbf{i}) can be simulated (black lines). }}
\end{figure*}

In this experiment, we fabricated Josephson trijunctions on the surface of Bi$_2$Te$_3$ flakes and used magnetic flux to control the phase differences in the junctions. Figure 1a and 1b are the scanning electron microscopic (SEM) image of such a device. The three superconducting Pb pads, separated by $\sim$560 nm, couple with each other through Bi$_2$Te$_3$ to form Josephson junctions. The phase differences across the junctions can be adjusted either simultaneously by applying a global magnetic field, or individually by applying local currents to the two half-turn coils. The Andreev bound states (ABSs) of the trijunction can be detected by measuring the contact resistance of the Au electrodes, which contact the Bi$_2$Te$_3$ surface at the center and the ends through windows on the blackish-looking insulating mask made of over-exposed polymethyl methacrylate (PMMA). For further information on device fabrication and measurement configuration please see the supplementary materials \cite{25}.

\begin{figure*}
\includegraphics[width=0.75 \linewidth]{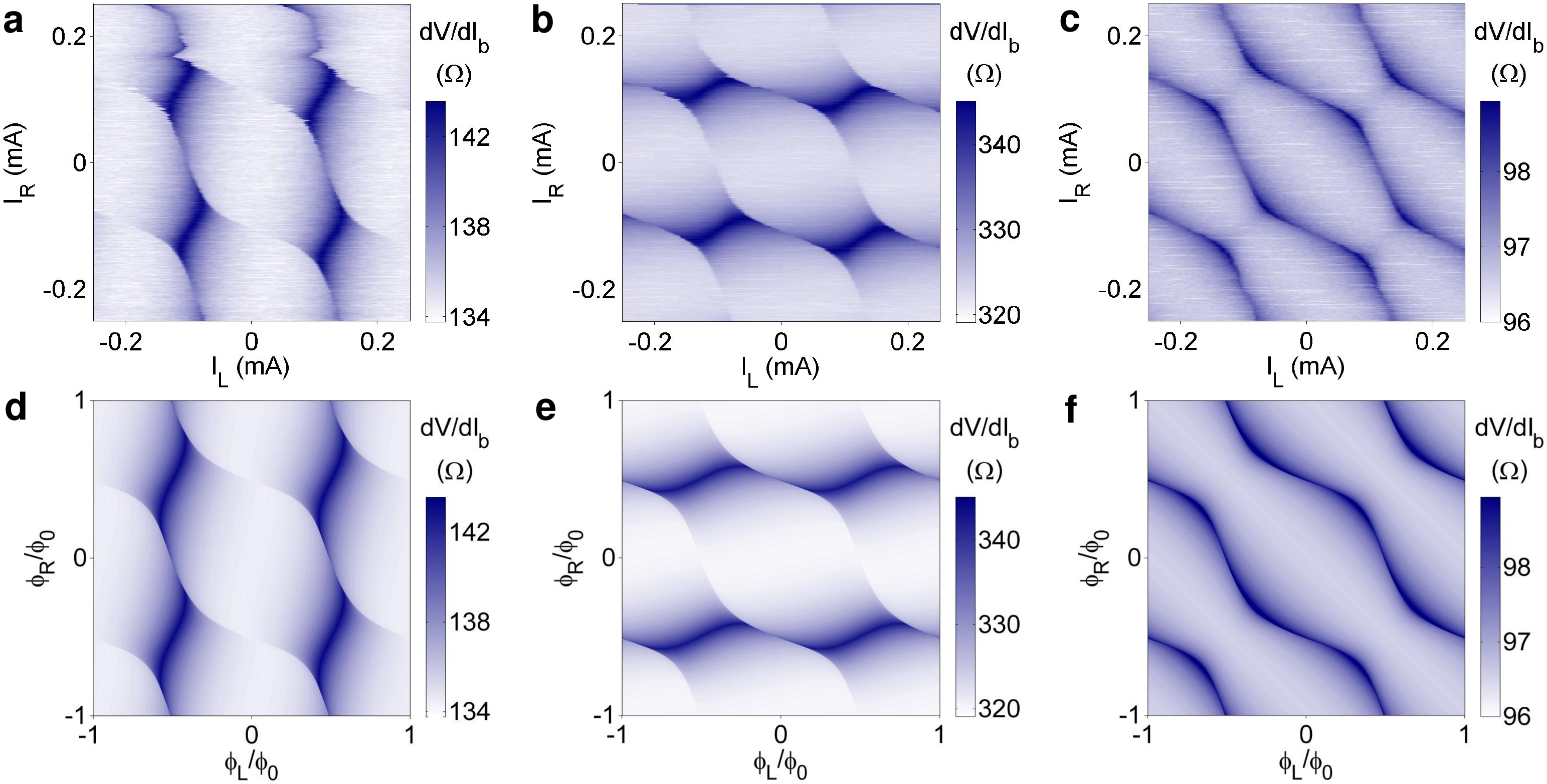}
\caption{\label{fig:fig3} {
The 2D data of $dV/dI_{\rm b}$ measured at (\textbf{a}) the left end, (\textbf{b}) the right end, and (\textbf{c}) the bottom end of the 1$^{\rm st}$ trijunction at $I_{\rm b}$=0, $T$=0.5 K, and by feeding currents $I_{\rm L}$, $I_{\rm R}$ to the half-turn coils to control the magnetic flux in the superconducting loops. (\textbf{d}), (\textbf{e}), (\textbf{f}) The simulated $dV/dI_{\rm b}$ at the left end, the right end, and the bottom end of the trijunction, respectively.}}
\end{figure*}

Let us first look at the data measured at the ends of the 1$^{\rm st}$ trijunction. Figure 2a shows the differential contact resistance $dV/dI_{\rm b}$ of the Au electrodes at the left end as functions of global magnetic field $B$ and bias current $I_{\rm b}$. Figure 2b shows the vertical line cuts of the data in Fig. 2a, namely the $dV/dI_{\rm b}~vs.~I_{\rm b}$ curves, at global magnetic fields indicated by the arrows with corresponding color in Fig. 2a. Figure 2c shows the horizontal line cut of the data, namely the $dV/dI_{\rm b}~vs.~B$ curve, at $I_{\rm b}$=0. Figure 2d, 2e, 2f, and Fig. 2g, 2h, 2i show similar data obtained at the right and the bottom ends of the trijunction, respectively. The measurements were performed at 0.5 K to avoid hysteresis (will be explained later).

We can see that the $dV/dI_{\rm b}$ at the left and the right ends shows similar behaviors. When the $dV/dI_{\rm b}~vs.~B$ curves in Fig. 2c and 2f enter into a low-resistance state, the $dV/dI_{\rm b}~vs.~ I_{\rm b}$ curves demonstrate a pronounced valley centering at zero bias (the red curves in Fig. 2b and 2e). When the $dV/dI_{\rm b}~vs.~B$ curves touch the normal-state value represented by the dashed lines in Fig. 2c and 2f, the $dV/dI_{\rm b}~vs.~I_{\rm b}$ curves become constant (the green curves in Fig. 2b, 2e). At the bottom end of the trijunction, differently, the low-resistance state at zero bias remains at most magnetic fields (Fig. 2g, 2i) --- the $dV/dI_{\rm b}$ approaches to the normal-state value only temporally during the field sweeping.

The following is our explanation for the observed phenomena. It is known from previous studies \cite{20} that the minigap in the junction can be modulated from open to closed by varying the phase difference $\varphi$ of the junction via \cite{5,26,27}: $\Delta = \Delta_0 |\cos(\varphi/2)|$ (where $\Delta_0$ is the induced gap). This minigap predominantly determines the contact resistance $dV/dI_{\rm b}$ of the Au-Bi$_2$Te$_3$ interface. When the interface is in the transparent regime, which is the case for the 1$^{\rm st}$ and the 2$^{\rm nd}$ (shown in Fig. 4) devices, the $dV/dI_{\rm b}$ will be reduced within the minigap, due to Andreev reflections between Au and the induced superconducting Bi$_2$Te$_3$. When the interface is in the tunneling regime, e.g., for the 3$^{\rm rd}$ and the 4$^{\rm th}$ devices (shown in Fig. 4 and in the supplementary materials), the $dV/dI_{\rm b}$ will be enhanced within the minigap. In both regimes, the $dV/dI_{\rm b}$ can be described by the Blonder-Tinkham-Klapwijk (BTK) theory \cite{28}.

Through fitting the zero-magnetic-field $dV/dI_{\rm b}~vs.~I_{\rm b}$ curves in Fig. 2b, 2e and 2h using the BTK theory (the black lines), the barrier parameters of the Au contacts as well as the minigap $\Delta_0$ beneath the contacts can be obtained --- for the left end $\Delta_{\rm 0L}=15 \mu$eV, the number of channel $N_{\rm L}$=134, the barrier strength $Z_{\rm L}$=0.843; for the right end $\Delta_{\rm 0R}=15 \mu$eV, $N_{\rm R}$=63, $Z_{\rm R}$=0.825; and for the bottom end $\Delta_{\rm 0B}=7.0 \mu$eV, $N_{\rm B}$=200, $Z_{\rm B}$=0.753. The details can be found in the supplementary materials \cite{25}.

An applied magnetic flux $\phi$ in the superconducting loop modifies the phase difference $\varphi$ across the junctions, hence modifies the minigap in the junctions. When the field-induced screening supercurrents in the loop is negligibly small, namely the screening parameter \cite{29}: $\beta=2\pi LI_{\rm c}\phi/\phi_0$ approaches to zero (where $L$ is the inductance of a single loop, $I_{\rm c}$ the bulk critical supercurrent \cite{20} of the single junction, and $\phi_0$ the flux quantum), we simply have $\varphi=2\pi\phi/\phi_0$, so $\Delta = \Delta_0 |\cos(\pi\phi/\phi_0)|$, as represented by the blue lines in Fig. 2k, 2l. Otherwise, when the screening supercurrent cannot be neglected, $\varphi=2\pi\phi_{\rm e}/\phi_0$, where the effective magnetic flux $\phi_{\rm e}$ obeys the relation $\phi_{\rm e}=\phi-(\beta\phi_0/2\pi)\sin(2\pi\phi_{\rm e}/\phi_0) -(\beta\phi_0/2\pi)\sin(4\pi\phi_{\rm e}/\phi_0)$. In particular, when $\beta$ exceeds 0.5 (instead of 1, since the total screening supercurrent in one loop is 2$I_{\rm c}$ in our devices), $\phi_{\rm e}$ becomes multi-valued, so that hysteresis occurs during backward and forward field sweepings. We did observe hysteretic responses of $dV/dI_{\rm b}$ at the base temperature. The results are shown in the supplementary materials \cite{25}.

\begin{figure*}
\includegraphics[width=0.95 \linewidth]{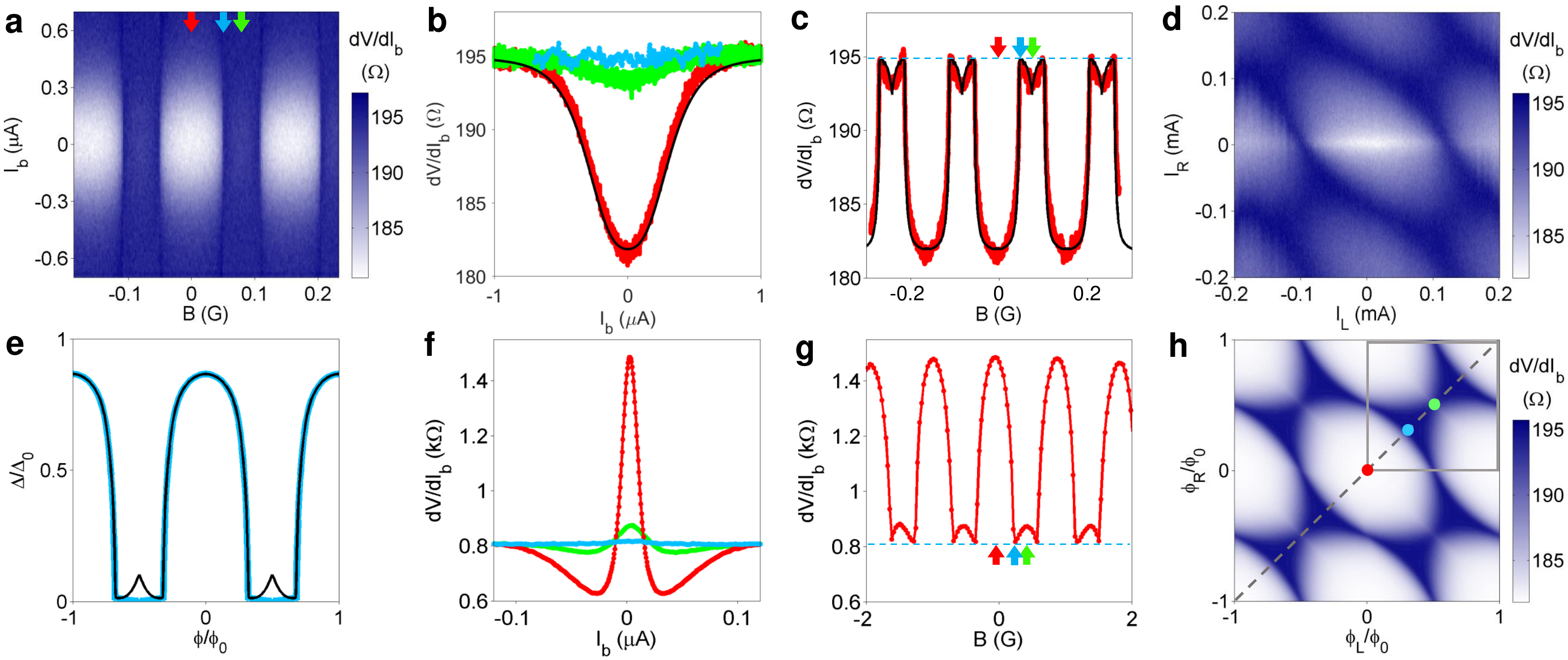}
\caption{\label{fig:fig4} {The $dV/dI_{\rm b}$ measured at the center of the trijunctions. (\textbf{a}) The $dV/dI_{\rm b}$ measured on the 2$^{\rm nd}$ trijunction at $T$=0.15 K, as functions of $I_{\rm b}$ and $B$ varying along the dashed line in (h). (\textbf{b}) The vertical line cuts of (a) at magnetic fields indicated by the arrows with corresponding color in (a). The data at $B$=0 are fitted by using the BTK theory (black lines). (\textbf{c}) The horizontal line cut of (a) at $I_{\rm b}$=0. (\textbf{d})/(\textbf{h}) The measured/simulated $dV/dI_{\rm b}$ at the center of the 2$^{\rm nd}$ trijunction in 2D flux space at $T$=0.25 K. The region enclosed by the gray square in (h) corresponds to Fu-Kane's MZM phase diagram shown in Fig. 1(d). (\textbf{e}) The minigap expected from the effective Hamiltonian via lattice-model numerical simulation, in the long junction limit (blue line) and with a finite junction length of 1.5 $\mu$m (black line). Using the functional form of this black line, the data in (c) can be simulated (black line). (\textbf{f}) and (\textbf{g}) The $dV/dI_{\rm b}~vs.~I_{\rm b}$ and $dV/dI_{\rm b}~vs.~B$ curves measured at the center of the 3$^{\rm rd}$ trijunction whose contact was in the tunneling regime. $T$=30 mK. }}
\end{figure*}

The data presented in Fig. 2 were collected at an elevated temperature of 0.5 K, at which the critical supercurrent was reduced such that the hysteretic behavior marginally disappeared, corresponding to the case of $\beta\approx$0.5. In this case, the effective magnetic flux $\phi_{\rm e}$ follows the warping line in Fig. 2j. As a result, the field dependence of the minigap in the left and right junctions (bottom junction) is modified to the black line in Fig. 2k (Fig. 2l).

With these field-dependent functional forms of minigap, together with the fitting parameters including $\Delta_0$ obtained from the $dV/dI_{\rm b}$~vs.~$I_{\rm b}$ curves, the global magnetic field dependences of $dV/dI_{\rm b}$ can be simulated by using the BTK theory. The results of simulation are shown as the black lines in Fig. 2c, 2f and 2i. Good agreements with the experimental data are obtained.

One noticeable feature in Fig. 2c, 2f and 2i is that the $dV/dI_{\rm b}$ approaches to and leaves away the normal-state value (the dashed lines) sharply, showing a linear closing of the minigap, intuitively hinting that the underlying mechanism of gap-closing and re-opening is a sign-change process originated from a 4$\pi$ periodicity, by which the complete closing of minigap (i.e., crossing of electron-like and hole-like ABSs) is guaranteed. We note that for trivial Josephson single junctions or trijunctions, the minigap will never oscillate to zero, due to unavoidable anti-crossing between electron-like and hole-like ABSs, even in highly transparent atomic point contacts \cite{30} or S-N-S junctions with N being a single graphene layer \cite{31}. The observation of full transparency in our Pb-Bi$_2$Te$_3$-Pb junctions, in which the two Pb electrodes are separated as far away as $\sim$560 nm, has to arise from a topologically protected mechanism based on MBSs, as has been studied previously \cite{20}.

Besides measuring the global magnetic field dependence of the $dV/dI_{\rm b}$, which explores along the diagonal direction in the two-dimensional (2D) flux space, we also measured the $dV/dI_{\rm b}$ over entire 2D flux space by individually adjusting the magnetic flux in the left and the right superconducting loops through the two half-turn coils. Shown in Fig. 3a, 3b, and 3c are the data acquired at the left, the right, and the bottom ends of the trijunction at $I_{\rm b}$=0, respectively. Figure 3d, 3e, and 3f are the simulated $dV/dI_{\rm b}$ at corresponding positions by using the minigap-phase relations and the BTK theory. Excellent agreements were obtained. We note that, because the data in Fig. 3a were measured in a different cooldown, the parameters for generating Fig. 3d are slightly different from those obtained in the above, being $\Delta_{\rm 0L}=13 \mu$eV, $N_{\rm L}$=149 and $Z_{\rm L}$=0.813.

From Fig. 3 we can see that the gap-opening and gap-closing regions form nearly straight stripes along the vertical, the horizontal, or 45$^o$ directions in the 2D flux space. It reflects that the minigap in the left/right junction is dominantly controlled by the magnetic flux in the left/right superconducting loop, and the minigap at the bottom junction is controlled by the magnetic flux in both loops. The slight warping of the stripes is due to the coupling of the two loops through the screening supercurrent flowing through the bottom junction. The warping should disappear in the $\beta\rightarrow$ 0 limit.

Let us now present the $dV/dI_{\rm b}$ measured at the center of the trijunctions. Due to malfunctioning of the central Au electrode of the 1$^{\rm st}$ device, the data were taken on the 2$^{\rm nd}$ and the 3$^{\rm rd}$ devices. Figure 4a shows the $dV/dI_{\rm b}$ measured on the 2$^{\rm nd}$ device (whose design is identical to the 1$^{\rm st}$ one) as functions of $B$ and $I_{\rm b}$ at an elevated temperature of 0.15 K (such that $\beta\approx$0.5). Figure 4b shows the vertical line cuts of Fig. 4a at three different fields indicated by the arrows in corresponding color. And Fig. 4c shows the horizontal line cut of Fig. 4a at $I_{\rm b}$=0. Also shown in Fig. 4f and 4g are the $dV/dI_{\rm b}~vs.~I_{\rm b}$ and $dV/dI_{\rm b}~vs.~B$ curves, respectively, measured on the 3$^{\rm rd}$ trijunction whose central contact was in the tunneling regime.

It can be seen that with sweeping global magnetic field along the diagonal direction of the flux space (i.e., along the dashed line in Fig. 4h), the minigap at the center of the trijunction varies periodically from open to closed to slightly re-opened. Complete gap-closing takes place near the edges in the bluish regions in Fig. 4h, as evident by the facts that the $dV/dI_{\rm b}$ there reaches the normal-state values represented by the horizontal dashed lines in Fig. 4c and 4g, in an accuracy of 100$\pm$3\% for the 2$^{\rm nd}$ device and 99$\pm$1.3\% for the 3$^{\rm rd}$ device. The way that the $dV/dI_{\rm b}~vs.~B$ curves touch the dashed lines is again in sharp peaks (i.e., linear closing), hinting that the underlying mechanism of gap-closing and re-opening is a sign-change process. Up on the gap-closing, the $dV/dI_{\rm b}~vs.~I_{\rm b}$ curves become completely flat (the blue curves in Fig. 4b, 4f). Such behaviors are impossible to arise from a trivial Josephson trijunction \cite{32}, in the latter a significant gap-like feature will remain on $dV/dI_{\rm b}~vs.~I_{\rm b}$ curves even if the transmission coefficient is as high as 0.9.

To understand why the minigap slightly re-opens between the two $dV/dI_{\rm b}$ peaks/dips in Fig. 4c/g, where complete gap-closing would be expected according to Fu-Kane's MZM phase diagram, we carried out numerical simulations based on the effective Hamiltonian of chiral Majorana states \cite{5,33}: $H_{\rm eff}=i\hbar v_{\rm M} (\gamma_l\partial_{\rm x}\gamma_l - \gamma_r\partial_{\rm x}\gamma_r) + i\delta\gamma_l\gamma_r$, where $\gamma_l$, $\gamma_r$ are the two counter-propagating chiral Majorana states in the junction, $\hbar$ the reduced Planck constant, $v_{\rm M}$ the effective group velocity of the chiral Majorana states, and $\delta=\Delta_0\cos(\varphi/2)$ is the coupling between the two states. The details can be found in the supplementary materials \cite{25}. For trijunctions in the long-length limit, we found that the global magnetic field dependence of minigap follows the blue line in Fig. 4e, supporting that the minigap closes completely in the entire shadowed regions of Fu-Kane's MZM phase diagram.

Our simulation also reveals that the boundary state at the center spreads slightly to the surrounding junctions. The spreading, hence the coupling to the surroundings, cannot be neglected when the length of the junction is finite, leading to the slight re-opening of the minigap. The spreading/re-opening becomes most significant at the vertexes of the shadowed regions, resulting in the small cusps at $\pm 0.5\phi_0$ on the black line in Fig. 4e. Using the functional form of the minigap represented by this black line, the global field dependence of $dV/dI_{\rm b}$ of the 2$^{\rm nd}$ trijunction can be simulated by using the BTK theory, with the parameters obtained through fitting the red curve in Fig. 4b: $\Delta_{B=0}=11.5 \mu$eV ($\Delta_0=13.3 \mu$eV), $N$=123, $Z$=0.931, and $T$=0.25 K. The result of simulation is shown as the black line in Fig. 4c.

By using the two half-turn coils, we further measured the zero-bias $dV/dI_{\rm b}$ of the 2$^{\rm nd}$ trijunction over the entire 2D flux space. The results are shown in Fig. 4d. Due to poor electrical connections to the right half-turn coil, the whole device heated up at high $I_{\rm R}$, which smeared out some of the details observed in global field sweeping at lower temperatures. Nevertheless, we can still see that the 2D data of $dV/dI_{\rm b}$ measured at the center is qualitatively different from those measured at the ends (Fig. 3), showing gap-closing over extended regions in phase space. With the same fitting parameters as above and the functional form of the minigap obtained from the lattice-model numerical simulation, the 2D data can be roughly simulated (Fig. 4h). The overall patterns of the measured and the simulated 2D data agree with each other, demonstrating the effectiveness of the MZM phase diagram predicted by Fu and Kane. Besides, both the measured and simulated patterns show that the edges of the MZM regions become curved in flux space, due to the same mechanisms (loop inductance and inter-loop coupling) that cause the warping of the patterns in Fig. 3.

To summarize, we have succeeded in fabricating Josephson trijunctions and controlling the phase differences in the trijunctions with the use of magnetic flux. We observed that the minigap at the center of the trijunctions undergoes complete closing near the edges of the shadowed regions in Fu-Kane's phase diagram, and gets slightly re-opened in the vicinity of the vertexes of the shadowed regions. We demonstrated through numerical simulation that such re-opening is a finite-size effect of the trijunctions. We also showed that the edges of the shadowed regions, near which the MZM appears at the center of the trijunctions, become curved in flux space when the loop inductance cannot be neglected. These findings provide the necessary details for further braiding MZMs by using sequences of magnetic flux pulses, towards the realization of surface code architectures \cite{23,24} and scalable TQC on TI-based two-dimensional platform.

\vspace {0.5 cm}

\noindent\textbf{Acknowledgments} We would like to thank L. Fu, F.C. Zhang, X.C. Xie, Q.F. Sun, X.J. Liu, X. Liu, G.M. Zhang, and Z.G. Cheng for fruitful discussions. This work was supported by the National Basic Research Program of China from the MOST grants 2016YFA0300601, 2017YFA0304700, and 2015CB921402, by the NSF China grants 11527806, 91221203, 11174357, 91421303, 11774405, and by the Strategic Priority Research Program B of the CAS grant No. XDB07010100.

\begin{widetext}


\section*{Supplementary material}

# Supplementary Materials for
# "Experimental realization of Josephson trijunctions for hosting Majorana zero modes"

Guang Yang[1,2,*], Zhaozheng Lyu[1,2,*], Junhua Wang[1,2], Jianghua Ying[1,2], Xiang Zhang[1,2], Jie Shen[1,#], Guangtong Liu[1], Jie Fan[1], Zhongqing Ji[1], Xiunian Jing[1], Fanming Qu[1,3,†] and Li Lu[1,2,3,†]

[1]*Beijing National Laboratory for Condensed Matter Physics, Institute of Physics, Chinese Academy of Sciences, Beijing 100190, China*
[2]*School of Physical Sciences, University of Chinese Academy of Sciences, Beijing 100049, China*
[3]*CAS Center for Excellence in Topological Quantum Computation, University of Chinese Academy of Sciences, Beijing 100190, China*

## Contents

1. Additional information on material characterization, device fabrication and measurement configuration
2. Additional data measured on more trijunction devices
3. Hysteretic behavior caused by the loop inductance
4. Fitting and simulating the data taken at the ends of the 1$^{st}$ trijunctions by using the BTK theory
5. Fitting and numerical simulating the data taken at the center of the 2$^{nd}$ trijunction by using the effective Hamiltonian for chiral Majorana states

---

[*] These authors contributed equally to this work.

[#] Present address: QuTech and Kavli Institute of Nanoscience, Delft University of Technology, 2600 GA Delft, The Netherlands.

[†] Corresponding authors, fanmingqu@iphy.ac.cn, lilu@iphy.ac.cn

1. **Additional information on material characterization, device fabrication and measurement configuration**

$Bi_2Te_3$ single crystals were grown by Bridgeman method and were confirmed to be of high quality by X-ray diffraction [1]. The carriers are of electron type, with a concentration of ~$2\times10^{18}$ cm$^{-3}$ at 2 K [1]. $Bi_2Te_3$ flakes were mechanically exfoliated from the bulk single crystals to Si/SiO$_2$ substrates. The thickness of the flakes was 100 nm or slightly less, being much thicker than a few monolayers so that the coupling between top and bottom surfaces can be neglected. On these flakes, the Pb trijunction as well as the superconducting loops were patterned by using electron beam lithography and deposited via magnetron sputtering. In addition, the two Al or Nb half-turn coils were deposited afterwards. Then, over-exposed PMMA was employed as an insulating mask, which covered the whole flake, the trijunction, and the two superconducting loops, except at the positions marked by yellow dots in Fig. 1c of the main manuscript where small windows of diameter 430 nm were open. Finally, Au electrodes were fabricated to contact the $Bi_2Te_3$ surface through the small windows for contact resistance measurement.

The contact resistance $dV/dI_b$ was measured by using a three-terminal configuration, in which the Au electrode whose contact resistance is to be measured is shared by the current injection loop and the voltage measurement loop, such that only the voltage drop across the contact is detected. Lock-in amplifiers were used to measure the differential resistance.

Previous studies reveal that the bulk carriers in our Pb-$Bi_2Te_3$-Pb junctions contribute the majority part of the supercurrent, giving rise to a Fraunhofer pattern in the diffusive transport limit [1, 2]. In such limit, it is well known that the minigap should not oscillate with magnetic flux. The observed $dV/dI_b$ oscillation thus has to arise from the oscillation of minigap in the surface states of $Bi_2Te_3$ [2]. This assignment is reasonable, since the contact resistance should depend most sensitively on the surface states.

## 2. Additional data measured on more trijunction devices

**2.1** A statistics and explanation on the data taken from the devices

The data presented in the main manuscript and in the Supplementary Materials below were obtained on four devices. Due to the technical difficulties - because many procedures are needed to fabricate the devices, only on the 2nd device we were able to take a complete set of data. But imperfections still remain for the 2nd device: one of the half-turn coils heated up during applying local flux. Nevertheless, after investigated a total of six trijunction devices and many single junction devices, we are fully confident on the reliability of the phenomena we found.

The following table summarizes the situation of data taking:

|  | End Contacts | | Central Contact | |
| --- | --- | --- | --- | --- |
|  | Flux along diagonal directions | 2D mapping | Flux along diagonal directions | 2D mapping |
| 1st device | Fig. 2 | Fig. 3 | Fall | Fall |
| 2nd device | SFig. 2.2 | SFig. 2.1 | Fig. 4 | Fig. 4 |
| 3rd device | SFig. 2.4 | N/A | Fig. 4, SFig. 2.4 | N/A |
| 4th device | SFig. 2.5 | N/A | SFig. 2.5 | N/A |

For the 1st device, everything worked perfectly except for the central contact. So, the junction states were successfully measured through the end contacts both along the diagonal direction by sweeping the background magnetic field, and in the entire 2D flux space by using the two half-turn coils, as shown in Fig. 2 and Fig. 3 in the main manuscript.

For the 2nd device, which is identical to the 1st one, all the contacts worked, but one of the half-turn coils heated up at large current. The measurements on the background field dependences of $dV/dI_b$ for all contacts were not influenced by the heating. The data for the central contact are presented in Fig. 4a, 4b and 4c, and the data for the end contacts are presented in SFig. 2.2 of the Supplementary Materials. The 2D mapping measurements were influenced by the heating. The 2D map for the central contact are presented in Fig. 4d, and the 2D map for the end contacts are presented in SFig. 2.1 of the Supplementary Materials.

For the 3rd and the 4th devices, which did not have half-turn coils, all contacts worked. The background field dependences of $dV/dI_b$ of all contacts are presented in SFig. 3.2 and SFig. 3.3 of the Supplementary Materials, and part of data are also presented in Fig. 4f and 4g of the main manuscript.

## 2.2 The $dV/dI_b$ measured at the ends of the 2nd trijunction

Due to malfunctioning of the central Au contact of the 1st trijunction device, the $dV/dI_b$ at the center of the trijunction, shown in Fig. 4 of the main manuscript, were taken on the 2nd device whose design was identical to the 1st one. Here we present the data of $dV/dI_b$ taken at the ends of the 2nd trijunction.

SFIG. 2.1 shows the 2D maps of $dV/dI_b$ measured at the ends of the 2nd trijunction at $I_b=0$ and T=0.15 K. The main features in Fig. 3 of the main manuscript are reproduced.

Also shown in the second row are the expected minigaps calculated according to the formulas presented in Section 4.2 of this Supplementary Information. The experimental data in the 2D maps can be directly compared with the minigap even without converting by using the BTK theory because the $dV/dI_b$ is in the tunneling regime.

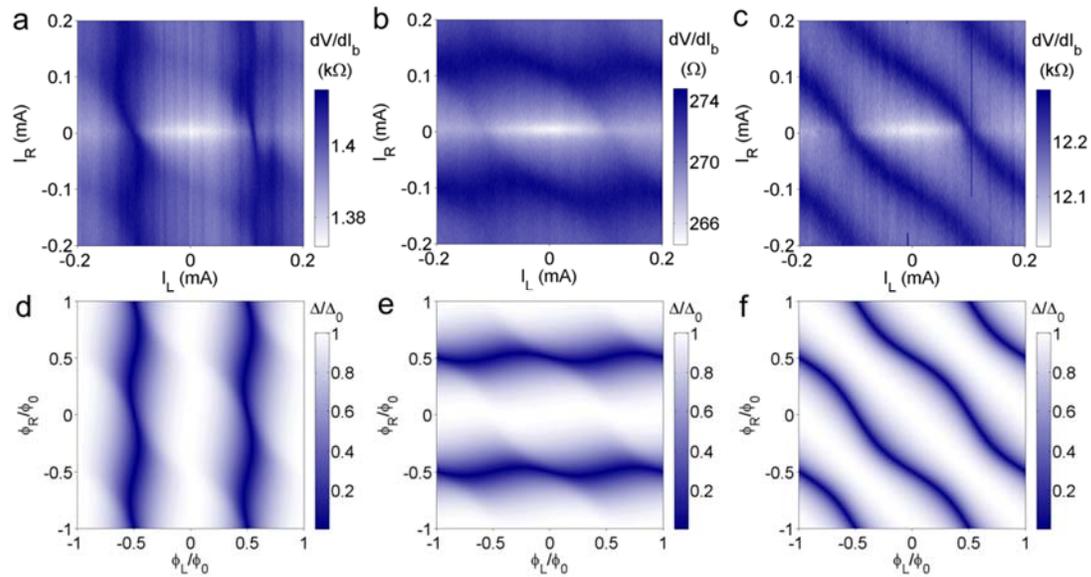

**SFIG. 2.1 | a, b, c,** The 2D maps of $dV/dI_b$ measured at the left end, the right end, and the bottom end of the 2nd trijunction at $I_b=0$, T=0.15 K. **d, e, f,** The expected minigaps at the left end, the right end and the bottom end of the 2nd trijunction, respectively, taking β=0.23.

SFIG. 2.2 shows the global magnetic field dependences of the $dV/dI_b$ measured at the left and the right end of the 2nd trijunction at $I_b=0$ and T=0.15 K. The Au contact at the bottom end became malfunctioning so that the data are unavailable. It can be seen that complete gap closing occurs, so that the valley structure in vertical line cuts totally disappears, and that the horizontal line cuts reach the normal-state values of $dV/dI_b$.

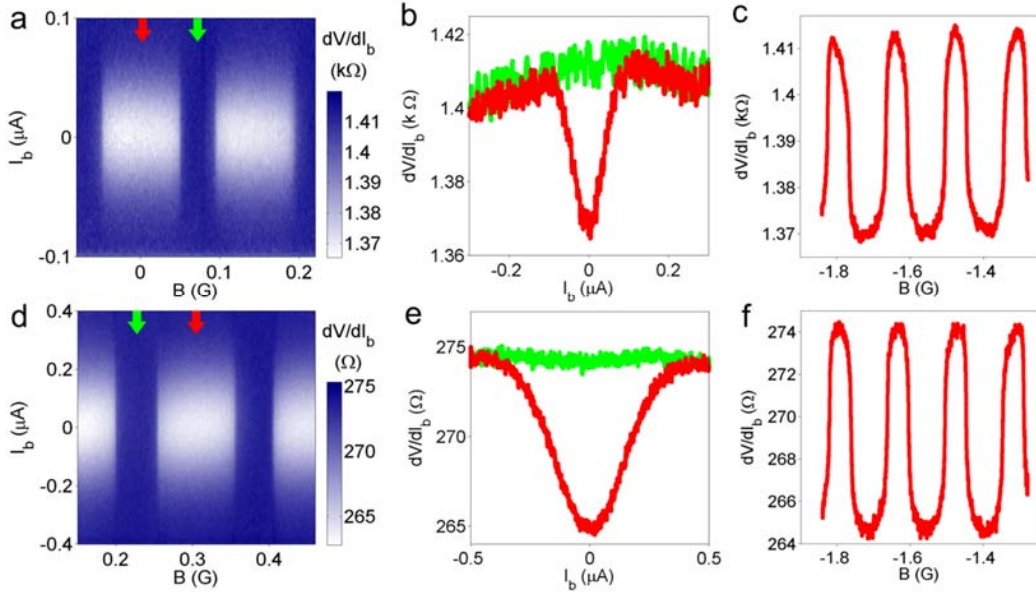

**SFIG. 2.2** | The $dV/dI_b$ measured at the left end (upper row) and the right end (lower row) of the 2$^{nd}$ trijunction, as functions global magnetic field and bias current. T=0.15 K. **b**, **e**, The vertical line cuts at magnetic fields indicated by the corresponding colors in (**a**) and (**d**). **c**, **f**, The horizontal line cuts at $I_b$=0.

**2.3** The 3$^{rd}$ and the 4$^{th}$ trijunctions with Pd contacts in the tunneling regime

For the 1$^{st}$ and the 2$^{nd}$ devices, the Au-Bi$_2$Te$_3$ interfaces for $dV/dI_b$ measurement were in the transparent regime. Here, we present the data taken on two more trijunction devices whose Pd-Bi$_2$Te$_3$ interfaces were in the tunneling regime, which was realized by properly controlling the device fabrication processes. We note that the contacting regime does not depend on the type of metals (Au or Pd) used, but depends on the fabrication processes.

Shown in SFIG. 2.3 are the SEM images of the 3$^{rd}$ device. The sizes of the Josephson junctions are about the same as in the 1$^{st}$ and the 2$^{nd}$ devices, but the shape of the superconducting loops is different. The loops of this device are significantly smaller than that in the 1$^{st}$ and the 2$^{nd}$ devices, so that the hysteretic behavior is absent. In addition, there were no half-turn coils for applying magnetic flux locally. So, with this device we could only apply a global magnetic field, to trace along the diagonal direction in Fu and Kane's MZM phase diagram.

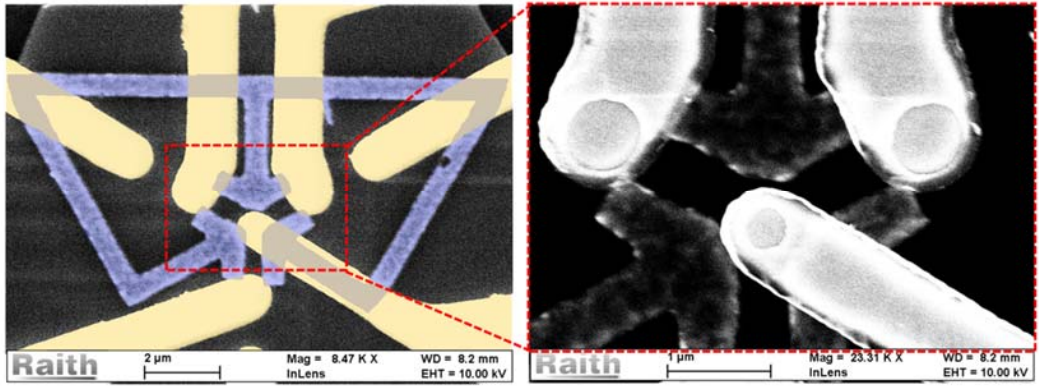

**SFIG. 2.3 |** SEM images of the 3$^{rd}$ device.

The measured data and the explanations can be found in SFIG. 2.4 and the caption therein.

Note that the dV/dI$_b$ measured at the center of this trijunction reaches the normal-state value cleanly, as shown by the dips touching the dashed line in SFIG. 2.4 k and l. This line shape, obtained in the tunneling regime, directly reflects the main feature of the simulated minigap shown in Fig. 4e of the main manuscript, being a property of the Majorana Hamiltonian [3, 4].

Moreover, when the dV/dI$_b$ approaches to the normal-state value, it demonstrates two sharp dips, which nicely corresponds to the two resistance peaks in the transparent regime, as found on the 2$^{nd}$ device and shown in Fig. 4 of the main manuscript. It represents that, the minigap closes sharply at the 45º edges of Fu and Kane's MZM phase diagram.

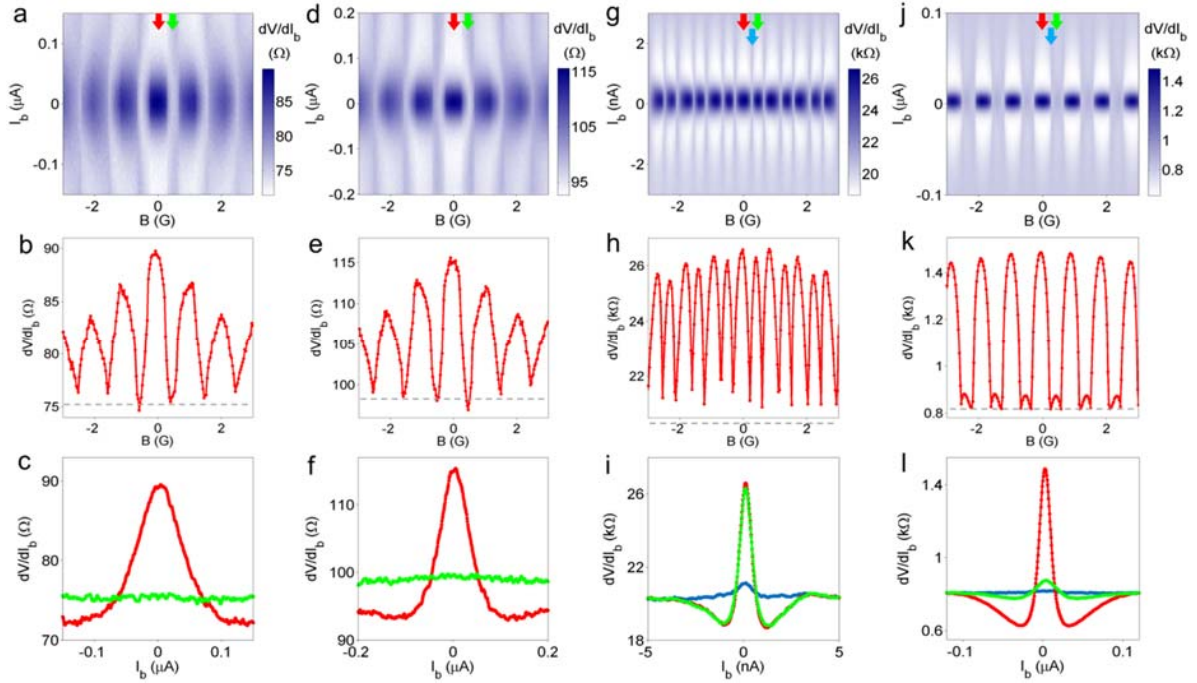

**SFIG. 2.4 |** The $dV/dI_b$ measured at the left end (first column), the right end (second column), the bottom end (third column), and the center (fourth column) of the 3$^{rd}$ trijunction, as functions of the global magnetic field and bias current, at T=30 mK. The second row shows the horizontal line cuts of the data in the first row at $I_b$=0. The dashed lines in (**b**), (**e**), (**h**) and (**k**) represent the normal-state values of $dV/dI_b$. The $dV/dI_b$ touches the dashed lines due to complete closing of the minigap. The third row shows the vertical line cuts at magnetic fields indicated by the colored arrows in the first row. Coherence peaks, here appeared as gross dips in $dV/dI_b$, can be seen. All these features can be readily fitted and simulated by using the BTK theory, as having been demonstrated previously [2].

SFIG. 2.5 shows the data obtained on the 4th trijunction device. This device has the same geometry and similar fabrication processes as for the 3rd one.

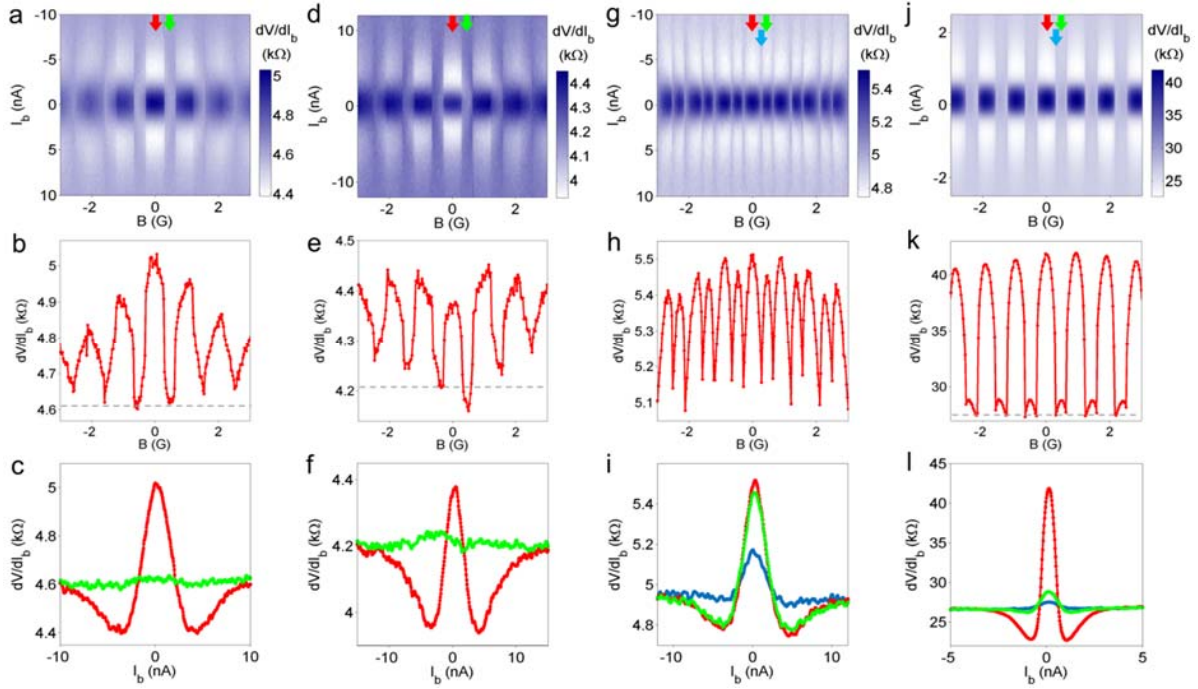

**SFIG. 2.5** | The $dV/dI_b$ measured at the left end (first column), the right end (second column), the bottom end (third column), and the center (fourth column) of the 4th trijunction, as functions of global magnetic field and bias current, at T=30 mK. The second row shows the horizontal line cuts of the data in the first row at $I_b$=0. The dashed lines in (**b**), (**e**), and (**k**) represent the normal-state values of $dV/dI_b$. The $dV/dI_b$ touches the dashed lines due to complete closing of the minigap. The third row shows the vertical line cuts at magnetic fields indicated by the colored arrows in the first row. Coherence peaks, here appeared as gross dips in $dV/dI_b$, can be seen. All these features can be readily fitted and simulated by using the BTK theory, as having been demonstrated previously[2].

## 3. Hysteretic behavior caused by the loop inductance

When the screening supercurrent in a SQUID loop is large enough, being able to modify the magnetic flux $\phi$ in the loop at the level of $\phi_0$ (the flux quantum), the effective magnetic flux $\phi_e$ becomes warping lines in SFIG. 3.1. As a result, the magnetic flux dependence of the minigap will be distorted from $\Delta=\Delta_0|\cos(\pi\phi/\phi_0)|$ to $\Delta=\Delta_0|\cos(\pi\phi_e/\phi_0)|$. For our trijunction during global magnetic field sweeping, $\phi_e$ at the left/right end is determined by the relation:

$$\phi_e=\phi+(\beta\phi_0/2\pi)\sin(2\pi\phi_e/\phi_0)+(\beta\phi_0/2\pi)\sin(4\pi\phi_e/\phi_0)$$

where $\beta=2\pi LI_c/\phi_0$ is the screen parameter of the superconducting loop [5], L is the loop inductance, and $I_c$ the half critical supercurrent of one loop in our device.

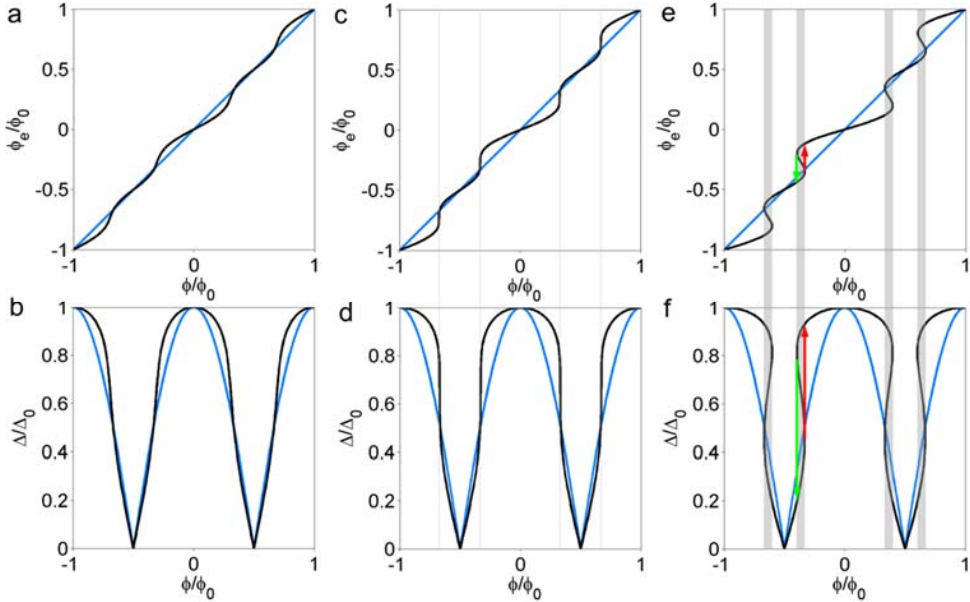

**SFIG. 3.1** | Warping effective magnetic flux and distorted minigaps at the left/right end of the trijunction. **a**, **b**, $\beta<0.5$. **c**, **d**, $\beta=0.5$. **e**, **f**, $\beta>0.5$. In this case, hysteresis appears in backward (green arrow) and forward (red arrow) field sweepings.

For our device, each superconducting loop contains two single Josephson junctions, the total critical supercurrent of a loop is $2I_c$, so that hysteresis is expected when $\beta>0.5$.

For the 1st device shown in the main manuscript, hysteretic behavior was indeed observed at low temperatures. The estimated $\beta$ at the base temperature of 30 mK is ~0.86 (see SFIG. 3.2). It corresponds to a critical supercurrent of $I_c\approx3.7$ μA. Note that the inductance L=76.6 pH can be obtained from the geometry of the loops.

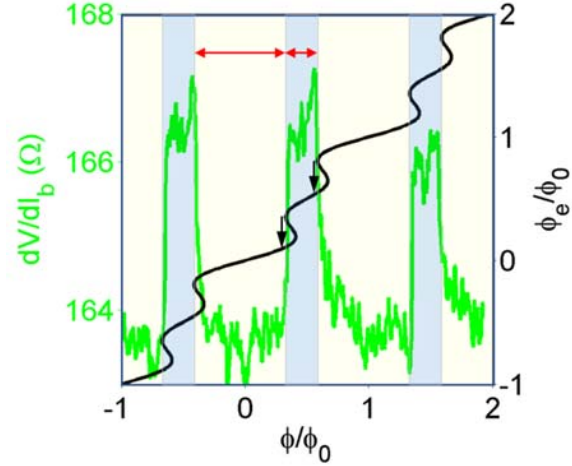

**SFIG. 3.2** | Estimating the β of the 1st trijunciton at the base temperature of 30 mK, from the widths of the peaks and the dips of the $dV/dI_b$. Different β yields different ratio between the two widths. The black line is for the case of β≈0.86, which fits to the width ratio of the data best.

By raising the temperature, $I_c$ (hence β) can be reduced, so that the hysteresis can be removed, as shown in SFIG. 3.3.

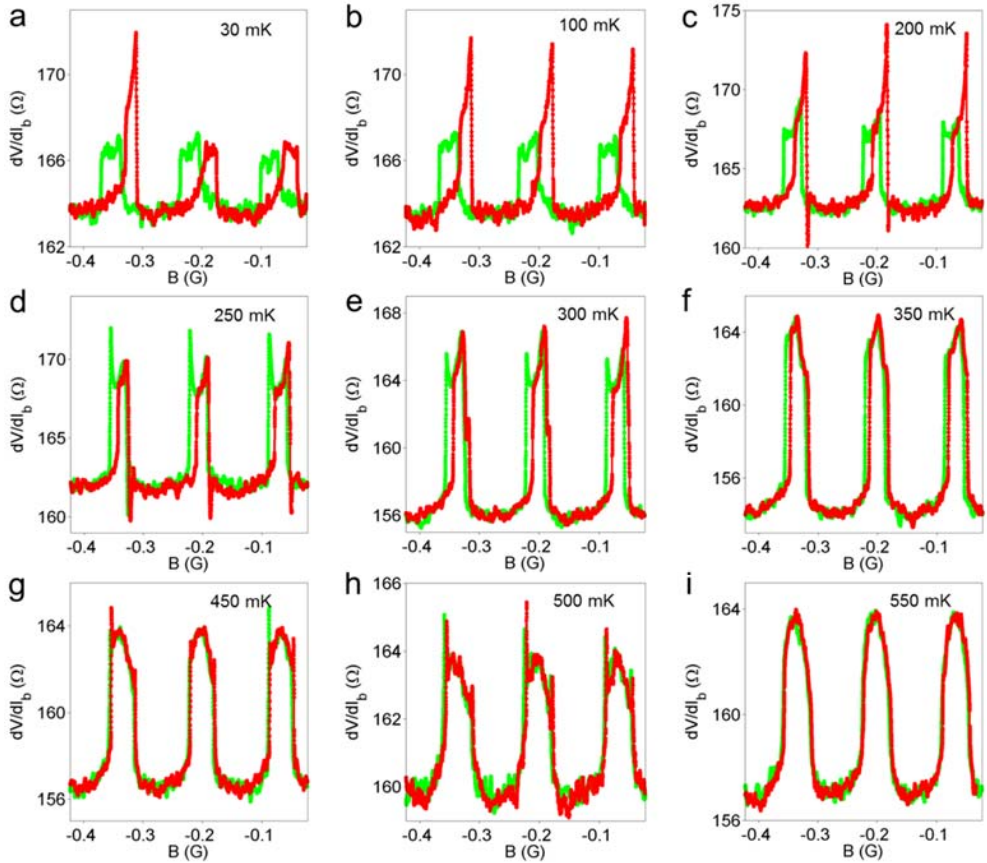

**SFIG. 3.3** | The temperature evolution of the hysteretic behavior seen from contact resistance measurements at the left end of the 1st trijunction in the main manuscript. The red traces were measured during ramping up, and the green ones during ramping down. Hysteresis marginally disappears at ~500 mK.

## 4. Fittings and simulating the data taken at the ends of the 1st trijunctions by using the BTK theory

Previously, we have successfully generalized the BTK theory to describe the single and two-particle processes across the interface between one metal and the second metal with Andreev bound states (ABS). Note that the BTK theory is applicable originally for the metal-superconductor interfaces [6]. In our generalization, the superconductor side is replaced with $Bi_2Te_3$ whose surface states contain induced ABS due to the lateral Andreev reflections between the two Pb-$Bi_2Te_3$ interfaces. The minigap between the electron-like and hole-like ABSs on the surface acts as a shutter, defining the effective energy window for current integration in the BTK treatment. Such a treatment successfully explains the observed phenomena in relevant devices [2].

For processing the data taken at the ends of the trijunctions, we first need to fit the $dV/dI_b$ vs. $I_b$ curves by using the BTK theory to obtain the minigap, than we use the amplitude of this minigap and the expected magnetic field dependence of the minigap, to further simulate the global magnetic field dependence of $dV/dI_b$, as well as to simulate the 2D maps of $dV/dI_b$ in the flux space.

According to the BTK theory, the total current across the Au-$Bi_2Te_3$ interface is [2, 6]:

$$I = \frac{2e^2}{h} N \int dE \left(1 - B(E) + A(E)\right)\left(f(E) - f(E - eV)\right)$$

where N is the number of conduction channels, e the electron charge, h the Planck constant, $f(E)=1/[1+\exp(E/k_BT)]$ is the Fermi distribution function, $k_B$ the Boltzmann constant, and A(E) and B(E) are the Andreev reflection coefficient and the normal-reflection coefficient, respectively.

A(E) and B(E) obey:

$$\left(1 - B(E) + A(E)\right) = \begin{cases} \dfrac{2\Delta^2}{(eV)^2 + (1 + 2Z^2)^2(\Delta^2 - (eV)^2)} & eV < \Delta \\ \dfrac{2eV}{eV + (1 + 2Z^2)^2\sqrt{((eV)^2 - \Delta^2)}} & eV > \Delta \end{cases}$$

where the dimensionless parameter Z is the barrier strength of the Au-$Bi_2Te_3$ interface, and $\Delta=\Delta_0|\cos(\varphi/2)|$ is the minigap of the junction [3, 7].

With the above formulas, we can calculate the $I_b$–V curve of the Au-$Bi_2Te_3$ interface, then to get the $dV/dI_b$ vs. $I_b$ curve. By fitting the calculated $dV/dI_b$ vs. $I_b$ curve to the measured data, the parameters $\Delta_0$, N,

and Z can be obtained at given temperature T.

**4.1** Fitting and simulating the data in Fig. 2 of the main manuscript

The black lines in Fig. 2b, e and h of the main manuscript are the fitted $dV/dI_b$ vs. $I_b$ curves, with fitting parameters
for the left end: $\Delta_{0L}=15$ µeV, $N_L=134$, $Z_L=0.843$, T=0.5K;
for the right end: $\Delta_{0R}=15$ µeV, $N_R=63$, $Z_R=0.825$, T=0.5K;
for the bottom end: $\Delta_{0B}=7.0$ µeV, $N_B=200$, $Z_B=0.753$, T=0.5K.

With the same fitting parameters and the functional forms of minigap shown as black lines in Fig. 2k, l of the main manuscript, the global magnetic field dependences of $dV/dI_b$ can be simulated by using the BTK theory with no additional parameters. The results are shown as black curves in Fig. 2c, f and i of the main manuscript.

**4.2** Fitting and simulating the data in Fig. 3 of the main manuscript

In order to simulate the 2D maps of $dV/dI_b$, we need the functional forms of $\Delta_L(\phi_L,\phi_R)$, $\Delta_R(\phi_L,\phi_R)$, and $\Delta_B(\phi_L,\phi_R)$.

For the left end: $\Delta_L(\phi_L,\phi_R)=\Delta_{0L}|\cos(\varphi_L/2)|=\Delta_{0L}|\cos(\pi\phi_{eL}/\phi_0)|$
For the right end: $\Delta_R(\phi_L,\phi_R)=\Delta_{0R}|\cos(\varphi_R/2)|=\Delta_{0R}|\cos(\pi\phi_{eR}/\phi_0)|$
For the bottom end: $\Delta_B(\phi_L,\phi_R)=\Delta_{0B}|\cos(\varphi_B/2)|=\Delta_{0B}|\cos[\pi(\phi_{eL}+\phi_{eR})/\phi_0]|$
where $\phi_L$, $\phi_R$ are the applied magnetic flux in the left and right loops, and $\phi_{eL}$, $\phi_{eR}$ are the effective magnetic flux in the left and right loops, respectively. They can be obtained from the following relations:
$\phi_{eL}=\phi_L-(\beta\phi_0/2\pi)\sin(2\pi\phi_{eL}/\phi_0)-(\beta\phi_0/2\pi)\sin(2\pi\phi_{eR}/\phi_0)$
$\phi_{eR}=\phi_R-(\beta\phi_0/2\pi)\sin(2\pi\phi_{eR}/\phi_0)-(\beta\phi_0/2\pi)\sin(2\pi\phi_{eL}/\phi_0)$

With the above functional forms, the 2D maps of $dV/dI_b$ at the right and the bottom ends in Fig. 3b, c of the main manuscript can be simulated by using the same fitting parameters as in Fig. 2, namely
for the right end: $\Delta_{0R}=15$ µeV, $N_R=63$, $Z_R=0.825$, T=0.5K;
for the bottom end: $\Delta_{0B}=7.0$ µeV, $N_B=200$, $Z_B=0.753$, T=0.5K.

Because the data in Fig. 3a were measured in a different cooldown, the parameters are slightly different from those used in Fig. 2a, being

$\Delta_{0L}$=13 µeV, $N_L$=149, $Z_L$=0.813 and T=0.5 K. These parameters are obtained by fitting the specially measured $dV/dI_b$ vs. $I_b$ curve in the second cooldown, as shown in SFIG. 4.1.

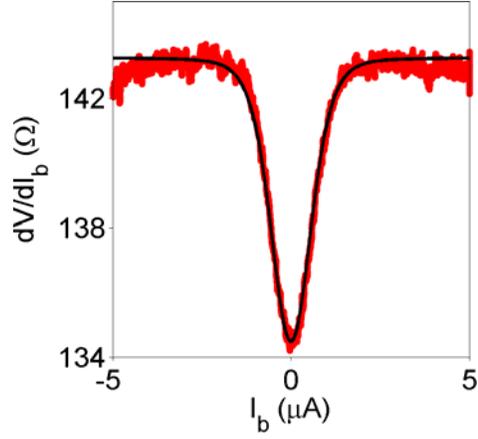

**SFIG. 4.1** | The $dV/dI_b$ vs. $I_b$ curve measured at the left end of the 1st trijunction in the second cooldown. The data are fitted by using the BTK theory (black line), through which the minigap $\Delta_{0L}$=13 µeV can be obtained. This value was further used to simulate the 2D map in Fig. 3a of the main manuscript.

## 5. Fitting and numerical simulating the data taken at the center of the 2nd trijunction by using the effective Hamiltonian for chiral Majorana states

**5.1** The lattice model

According to the theory of Fu and Kane [3, 4], the effective Hamiltonian for single Josephson junction constructed on the surface of topological insulator contains the kinetic energy of the two chiral Majorana states in the junction and the coupling energy between them:

$$H = -iv_M \hbar \sigma_z \partial_x + \delta(\varphi)\sigma_y$$

where $v_M$ is the effective group velocity, $\hbar$ is the reduced Planck constant, $\sigma_z$ and $\sigma_y$ are the Pauli matrix, $\delta(\varphi) = \Delta_0 \cos(\frac{\varphi}{2})$ is the mingap, and $\varphi$ is the phase difference of the junction. By solving this Hamiltonian, we can get the information of the ground and excited states of the trijunction, including the energies and the spatial distributions of these states.

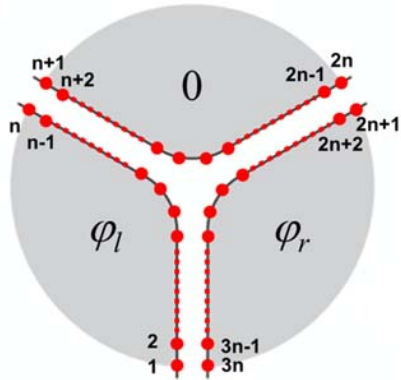

**SFIG. 5.1 | Schematic of the lattice model for the trijunction.** A trijunction contains three Josephson junctions formed of three superconducting pads on the surface of a topological insulator. The phases of the superconducting pads are labeled on the pad. The red dots are the discretized lattice points in the lattice model.

Fu and Kane have studied the trijunction in the long-length limit. For real trijunctions with finite size, it is difficult to solve the effective Hamiltonian analytically. We therefore try to discretize the system and simulate the solution of the Hamiltonian numerically based on the lattice model. The discretization is shown in SFIG. 5.1. The Hamiltonian is $H=T+V$, where $T$ is the kinetic term and $V$ the coupling term. The corresponding matrixes of $T$ and $V$ are shown below.

$$T = \begin{bmatrix} \begin{matrix} 0 & -it & 0 & 0 \\ it & \ddots & \ddots & 0 \\ 0 & \ddots & \ddots & -it \\ 0 & 0 & it & 0 \end{matrix} & 0_{n\times n} & 0_{n\times n} \\ 0_{n\times n} & \begin{matrix} 0 & -it & 0 & 0 \\ it & \ddots & \ddots & 0 \\ 0 & \ddots & \ddots & -it \\ 0 & 0 & it & 0 \end{matrix} & 0_{n\times n} \\ 0_{n\times n} & 0_{n\times n} & \begin{matrix} 0 & -it & 0 & 0 \\ it & \ddots & \ddots & 0 \\ 0 & \ddots & \ddots & -it \\ 0 & 0 & it & 0 \end{matrix} \end{bmatrix}$$

$$V = \begin{bmatrix} 0_{n/2\times n/2} & i\delta(\varphi_l)J_{n/2\times n/2} & & 0_{n/2\times n/2} & -i\delta(\varphi_l-\varphi_r)J_{n/2\times n/2} \\ -i\delta(\varphi_l)J_{n/2\times n/2} & 0_{n/2\times n/2} & 0_{n\times n} & 0_{n/2\times n/2} & 0_{n/2\times n/2} \\ & 0_{n\times n} & & 0_{n\times n} & & 0_{n\times n} \\ 0_{n/2\times n/2} & 0_{n/2\times n/2} & & 0_{n/2\times n/2} & i\delta(\varphi_r)J_{n/2\times n/2} \\ i\delta(\varphi_l-\varphi_r)J_{n/2\times n/2} & 0_{n/2\times n/2} & 0_{n\times n} & -i\delta(\varphi_r)J_{n/2\times n/2} & 0_{n/2\times n/2} \end{bmatrix}$$

where $t = v_M\hbar/(2a)$ is the hoping energy, $a$ is the space between two lattice points, $I$ the identity matrix, $J$ the reverse identity matrix, $\varphi_l$ and $\varphi_r$ are the phases of the left and right superconducting pads, respectively. In our device, when a global magnetic field $B$ is applied, $\varphi_l = -\varphi_r = 2\pi\phi/\phi_0$, where $\phi$ is the magnetic flux in one superconducting loop.

**SFIG. 5.2 | The energies and the spatial distributions of the states in the trijunction in the long-junction limit. a**, The flux dependence of the energies of the states, forming four groups. The blue curves represent the ground-state energy of the system. The spatial distribution of this state is localized at the center of the trijunction at most flux, as revealed by the spatial probability amplitude (PA) of this state in the bottom junction (**b**) and in the left and the right junctions (**c**). The red curve represents the lowest energy of the group of states in the bottom junction. And the cyan and the black-dash curves represent two degenerated lowest energies of the groups for the left and the right junctions. The parameters used in the simulations are: $t$=1.33 meV, $\Delta_0$=13.3 μeV, and the effective junction length is 8.9 μm (already in the long-junction limit). **b** and **c**, the PA of the system's ground state in the bottom junction and in the left and the right junctions, respectively, as functions of magnetic flux and position. 0 denotes the center and 1 denotes the ends of the trijunction.

The results of numerical solution for a trijunction in the long-junction limit are shown in SFIG. 5.2.

SFIG. 5.2a shows the energies of the ground and the excited states of the system at different flux. We can see that these energies/states form four different groups. The lowest energy of one of the groups is plotted as the red curve in SFIG. 5.2a. It follows approximately the same flux dependence as the minigap in the bottom junction: $E = \cos(2\pi\phi/\phi_0)$ (represented by the blue curve in Fig. 2l of the main manuscript). The calculation of the spatial probability amplitude (PA) of these states also confirms that they are distributed in the bottom junction. We call these states the junction states.

There are other two groups of junction states, with degenerated lowest energies plotted in SFIG. 5.2a as the cyan curve and the black-dash curve. The lowest energies of these junction states follow approximately the same flux dependence as the minigaps in the left and the right junctions: $E = \cos(\pi\phi/\phi_0)$ (represented by the blue curve in Fig. 2k of the main manuscript). These states are distributed in the left and the right junctions.

The fourth group is the state with energy represented by the blue curves in SFIG. 5.2a. Compared to the "bulk" states (i.e., the junction states of the other three groups), the state in the fourth group is an "edge" state, localized mostly at the center of the trijunction. This can be seen from the PA of this state (SFIG. 5.2b and 5.2c). With increasing flux from 0 to $\phi_0/8$, this state is the ground state of the whole system, so the PA is located at the center of the trijunction. From $\phi_0/8$ to $\phi_0/4$, the bottom junction seems to have the lowest energy, so the PA transfers to the bottom junction as shown in SFIG. 5.2b. Above $\phi_0/4$, the state of the fourth group keeps to be the ground state of the system, and is localized at the center of the trijunction.

From SFIG. 5.2a, when the flux is less than $\phi_0/4$, the phase difference in all three junctions are less than $\pi$, so that the gap in the junctions, between the electron-like and hole-like ABSs, are all positive. This leads to a positive minigap in the system's ground state. When the flux exceeds $\phi_0/4$, the phase difference of the bottom junction is larger than $\pi$, so that the $4\pi$-period electron-like and hole-like ABSs cross with each other, the minigap in this junction reverses. This leads to the appearance of a "positive gap" - "negative gap" boundary at the center of the trijunction, so that the minigap in system's ground state is closed locally, and that MZM appeared at the center of the trijunction.

From SFIG. 5.2b and 5.2c it can be seen that, the localized state at the center of the trijunction spreads to the surroundings slightly. The spreading to the bottom junction is significant at $\phi_0/4$, because the gap in the bottom junction is almost zero then. Similarly, the spreading to the left and the right junctions is significant at $\phi_0/2$, because the gap in these two junctions is almost zero then. Even within the range of $\phi_0/4$ to $\phi_0/2$, the spreading can still be seen, though very tiny. When the length of the junction is finite, such spreading couples the zero-energy state at the center to the states in the surroundings, as well as to the states at the ends, if any, resulting in the slightly re-opening of the minigap.

**5.2** The choice of parameters for simulating the data

To simulate the experimental data, we set the lattice spacing $a$ in the discretization model to 1 nm, and let the effective maximum wave vector to be $\pi/a$, which is about 0.3Å$^{-1}$, larger than 0.2 Å$^{-1}$, the Fermi wave vector of $Bi_2Te_3$. The length of junction is about 1.5 μm in our device, so that the number of lattice points on one of the chiral edge is $n=3000$.

SFIG. 5.3 shows the simulated minigap at different hopping energy $t$. By choosing $t=1.33$ meV, and with the amplitude of minigap $\Delta_0=13.3$ μeV obtained from fitting the red curve in Fig. 4b of the main manuscript, we can simulate the measured $dV/dI_b$ vs. B curve as shown in the right panel of SFIG. 5.3. The other parameters used in the simulation are: the number of channels $N=123$, the barrier strength $Z=0.931$, and the screening parameter $\beta=0.43$.

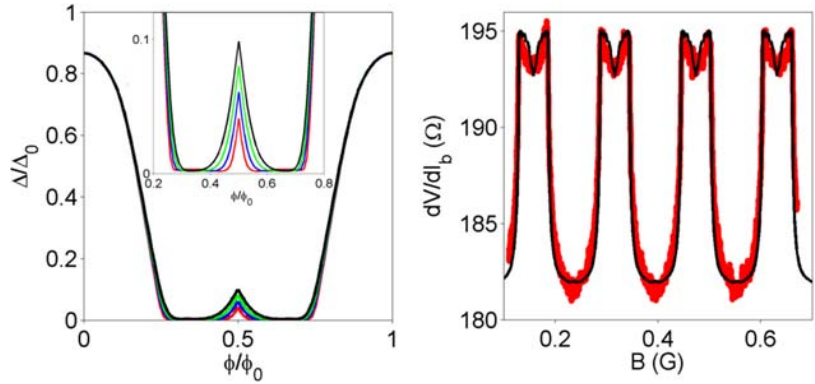

**SFIG. 5.3** | (left panel) The simulated minigap for trijunction with a finite length of 1.5 μm, at different hopping energy $t$ =0.532 meV (red), 0.798 meV (blue), 1.064 meV (green), 1.33 meV (black). This black curve, also shown in Fig. 4e of the main manuscript, is used to simulate the experiment data (left panel, the same as Fig. 4c in the main manuscript).

We note that the choice of $t$=1.33 meV implies that the effective group velocity of the chiral Majorana states is $v_M \approx 4\times 10^3$ m/s. Such a $v_M$ is considered reasonable [4].


\end{widetext}

\end{document}